\documentstyle[11pt,aas2pp4]{article}
\lefthead{Gabel et al}
\righthead{Photoionization Modeling of NGC 1052}
\begin{document}
\title{{\it HST} Observations and Photoionization Modeling of the LINER Galaxy NGC~1052\altaffilmark{1}}
\author{J. R. Gabel, F. C. Bruhweiler, D. M. Crenshaw, S. B. Kraemer, and C. L. Miskey}
\affil{Institute for Astrophysics and Computational Sciences} 
\affil{Department of Physics, The Catholic University of America, Washington, DC 20064}
\altaffiltext{1}{Based on observations made with the NASA/ESA {\it Hubble Space Telescope}, obtained from the data archive at the Space Telescope Science Institute. STScI is operated by the Association of Universities for Research in Astronomy, Inc. under NASA contract NAS~5-26555.}

\begin{abstract} We present a study of available {\it Hubble Space Telescope (HST)} spectroscopic and imaging observations of the
low ionization nuclear emission line region (LINER) galaxy NGC~1052.  The
WFPC2 imagery clearly differentiates extended nebular H$\alpha$ emission from that of the compact core.  Faint Object Spectrograph (FOS) observations provide a full
set of optical and UV data (1200-6800 $\AA$).  These spectral data sample the
innermost region (0."86~$\times$~0."86~$\sim$~82pc~$\times$~82pc) and exclude
the extended H$\alpha$ emission seen in the WFPC2 image.  The derived emission line 
fluxes allow a detailed analysis of the physical conditions within the nucleus.
The measured flux ratio for H$\alpha$/H$\beta$, 
F$_{H\alpha}$/F$_{H\beta}$~=~4.53, indicates substantial intrinsic reddening, 
E(B-V)=0.42, for the nuclear nebular emission.  This is the first finding of a large 
extinction of the nuclear emission line fluxes in NGC~1052. If the central ionizing continuum is assumed to be attenuated by a comparable amount, then the emission line
fluxes can be reproduced well by a simple photoionization model using a central power law continuum source with a spectral index of $\alpha$~=~$-$1.2 as deduced from the observed flux distribution. A multi-density, dusty gas gives the best fit to the observed emission line spectrum.  Our calculations show that the small contribution from a highly ionized gas observed in NGC~1052 can also be reproduced solely by photoionization modeling.  The high gas covering factor determined from our model is consistent with the assumption that our line of sight to the central engine is obscured.    
\end{abstract}               


\section{Introduction}      

     Low ionization nuclear emission line regions, (LINERs) have been found to occur
in nearly 1/3 of all bright, nearby galaxies (\cite{ho1995}).  They share many of the 
emission characteristics of
active galactic nuclei (AGN), but are distinguished by their low luminosities and relatively
strong, low-ionization emission lines (e.g., [O~I]~$\lambda$$\lambda$6300, 6363,
[O~II]~$\lambda$3727, [N~II]~$\lambda$$\lambda$6548,6584, [S~II] $\lambda$$\lambda$6716,6731).
 A recent study by Ho (1999) has demonstrated that the UV $-$ X-ray spectral energy distribution of LINERs is harder than in higher luminosity AGNs.  This is consistent with the general trend exhibited in AGNs of decreasing
$\alpha_{ox}$ with decreasing luminosity (\cite{avni1982}). The physical relation between LINERs and AGNs is unclear. They may
be simply the low luminosity extension of the AGN phenomenon, powered by accretion
onto a compact object or, alternatively, they may be the result of entirely different
mechanisms.

    Although central source photoionization appears to be the dominant mechanism powering the nebular emission seen in the high luminosity Seyfert galaxies, it has not been thoroughly tested for galaxies containing LINERs.  Competing ionization mechanisms that have been used to successfully explain at least some of the emission characteristics of LINERs include photoionization by hot
stars (\cite{terl1985,fili1992}), photoionization by non-thermal emission typical of luminous AGN (\cite{ferl1983,ho1993}) and
shock-heating (\cite{heck1980,dopi1995,dopi1996}). 
LINERs might represent a heterogeneous class of objects
since the proposed ionization mechanisms can all explain the low apparent luminosity of LINERs. Perhaps more than one of these various mechanisms is at work in the same object.

    NGC~1052 is a nearby (z=0.0045; \cite{deva1991}) elliptical galaxy (E4) with a strong  nuclear
emission line spectrum dominated by low ionization features.  For example, the
[O~I]~$\lambda$6300 and [O~II]~$\lambda$3727 fluxes exceed that of
[O~III]~$\lambda$5007, plus [N~II]~$\lambda$6584/ H$\alpha>$1 (\cite{fosb1978,ho1993,ho1997}).  Fosbury et~al.\ 1978, hereafter F78, showed that the emission line gas
is comprised of a luminous central component (d~$\leq$~2$\arcsec$) and a diffuse extended
region (d$\sim$20$\arcsec$).  Each emission region contributes approximately equally
in total Balmer line flux.  Furthermore, the extended component emits many of the
other prominent optical lines that are observed in the spectrum of the central region, such
as [O~II]~$\lambda$3727 and [O~III]~$\lambda$5007 (F78).  

         The observed optical continuum emission is dominated by the host galaxy, with
no clear detection of any underlying non-thermal emission (F78; \cite{fosb1981}).
 Radio studies, on the other hand, clearly show it to have both compact and extended
continuum emission components.  At $\sim$1.5 GHz, its radio emission is characterized
by a flat, unresolved ($\leq$~1$\arcsec$) core and two weaker lobes extending
10-15$\arcsec$ on either side of the compact source (\cite{wrob1984}).  NGC~1052 has been shown to have an IR excess in the 5-20~$\mu$m band within
the inner 2$\arcsec$ of the nucleus (\cite{beck1982}).   A recent analysis of {\it
ROSAT}/HRI imagery shows some extended soft X-ray emission, but the low spatial
resolution at these energies prevents isolating any compact emission at scales less
than 10$\arcsec$ (\cite{guai1999}). X-ray observations with {\it ASCA} indicate
a very flat spectrum for the hard X-ray emission (2-10~keV) of NGC~1052 (\cite{guai1999}).   

      As with LINERs in general, there is much disagreement on the nature of the
excitation mechanism responsible for the nuclear activity in NGC~1052.  Initial
findings have suggested that shock-heating could produce the observed line emission (\cite{kosk1976}; F78). 
This claim is largely based upon the high electron temperature (T$\sim$33,000 K)
implied by the strong  [O~III]~$\lambda$4363 emission.   Furthermore, simple shock
excitation models have been shown to match the prominent optical emission features 
(\cite{kosk1976}; F78). In an alternate interpretation, Ferland~\& Netzer~(1983), and more recently Ho et~al.~(1993), have shown that many of the strong emission lines observed in NGC~1052 (and in
LINERs in general) could be produced by photoionization from a nonthermal continuum
typical of Seyfert galaxies, namely a power law with spectral index, $\alpha$$\approx$-1.5.
The major difference in the models between the typically more luminous Seyferts and
the LINERs is that the ionizing flux is inferred to be significantly weaker in
LINERs.  Subsequent detailed photoionization modeling of NGC~1052 by P\'equinot (1984),
hereafter P84, has shown that many of the optical emission lines, including
[O~III]~$\lambda$4363, could be fit well by photoionization of a multi-density
component gas.  His study assumed a central photoionizing source consisting of a
black body continuum with an X-ray tail extending to higher energies.
Recently, a polarimetric study has shown NGC~1052 to have a broad H$\alpha$
emission line component with FWHM$\sim$5000~km~s$^{-1}$ (\cite{bart1998}).  This provides compelling evidence
for the presence of a compact object and lends strong support to the
non-stellar photoionization hypothesis.

      In this paper, we present a detailed
analysis of the nuclear activity in the prototypical LINER galaxy NGC~1052. 
Specifically, we use {\it Hubble Space Telescope (HST)} observations of NGC~1052 to explore if photoionization
represents a viable option in explaining the emission line spectrum in this LINER.
In Section~2 we
describe the HST imagery and spectroscopic
observations used in this study and present our measured emission line fluxes and an
analysis of the WFPC2 H$\alpha$ imagery. 
We examine the possible role of extinction on the emission line fluxes and ionizing continuum.  The methodology and
results of our photoionization modeling are given in Section~3.  In Section~4,
we discuss the implications of our results and compare them with previous
studies to form a comprehensive picture of the activity of NGC~1052.  Finally, 
we summarize our results and present our conclusions in Section~5. 

\section{Observations and Emission Line Measurements}

\subsection {FOS Observations}

    The spectral data used in this study were obtained with the Faint Object
Spectrograph aboard the {\it Hubble Space Telescope} ({\it HST}/ FOS) on
1997~January~13.  They were retrieved from the Space Telescope Science Institute
(STScI) data archives.  All observations were of the active nucleus of NGC~1052
through the 0$\arcsec$.86-pair square aperture (post-COSTAR), covering
1200-6800~$\AA$.  Thus, the aperture sampled the central 82~pc~$\times$~82~pc of
the nuclear region of NGC~1052, assuming D~=~19.6~Mpc for
H$_o$~=~75~km~s$^{-1}$~Mpc$^{-1}$.  The optical and near-UV data were obtained
with high resolution gratings giving $\lambda$/$\Delta\lambda$$\sim$1300, while
the UV spectrum has a lower resolution given by
$\lambda$/$\Delta\lambda$$\sim$250.  The individual spectra are described in
Table~1 along with their exposure times, grating settings, and spectral resolutions.  Because of a
guide-star lock failure, the data may have slightly degraded spectral resolution
and sample a slightly larger spatial region than the aperture size would
indicate.  If the extended regions emit differently than the nucleus, this may
have an effect on the observed emission line ratios and absolute fluxes. 

      Using the maps of Burstein \& Heiles (1982), we find a foreground Galactic
reddening value of E(B-V)~=~0.02 in the direction of NGC~1052.  Figure~1 shows
the spectrum in the rest frame of NGC~1052, corrected for Galactic extinction
using the extinction law of Savage~\& Mathis (1979). 

\subsubsection {UV Emission}

      Examination of the UV spectrum below 2000~$\AA$ in Figure~1 shows a weak
UV continuum compared to the optical continuum, although, there appears to be a
slight rise toward lower wavelengths.  This UV continuum shows no evidence for
recent starburst activity in the nucleus of NGC~1052.  Specifically, the common
signatures of hot O~and~B stars, namely broad wind features from C~IV~$\lambda$1549, Si~IV~$\lambda$1400, and N~V~$\lambda$1240 and
narrower photospheric lines (Neubig~\& Bruhweiler~1997, 1999), are absent within
the low detection limits of the data. Although the results are not definitive,
the absence of stellar features is consistent with the UV continuum being
produced by a non-thermal compact source. Numerous UV emission lines are
detected in the FOS spectrum, compared to the IUE study in which only two
lines, C~III]~$\lambda$1909 and C~II]~$\lambda$2326, were unambiguously measured
(\cite{fosb1981}).

\subsubsection {Emission-Line Measurements}

      Visual inspection of the optical and near-UV data at
$\lambda$~$>$~2000~$\AA$ in Figure~1 clearly shows that the host elliptical
galaxy dominates the continuum.  Strong stellar absorption features are evident
and may affect flux measurements of certain emission lines, especially the
Balmer lines.  In addition, small wavelength variations in the composite stellar
spectrum of the host galaxy may mimic emission line features.  To remove the
spectral contribution of the host galaxy, we have used template spectra of
elliptical galaxies from the {\it HST}/FOS archival database to approximate its
integrated stellar spectrum.  We selected a proper template spectrum for each of
the three optical gratings (G270H, G400H, and G570H) based upon the following
criteria: \\
\noindent 1) We required the prominent absorption features that do not coincide
with emission features in NGC~1052 to match those in the host spectrum of
NGC~1052 (namely, Mg~I~$\lambda$2852, Ca~II~K~$\lambda$3934,
G~Band~$\lambda$4301, Fe~I~$\lambda$4440, Fe~I~$\lambda$4532,
Mg~Ib~$\lambda$5176, Fe~I~$\lambda$5333, Ca~I/ Mg~H~$\lambda$5590, Na~I
Doublet~$\lambda$5893, and Ca~I/ Fe~I $\lambda$6161; \cite{prit1976,cren1985});\\
\noindent 2)  We required the overall galactic flux distribution to be similar
to that in NGC~1052.\\
The template spectra were then scaled in magnitude, so that the strengths of the
absorption features matched that of NGC~1052, and subtracted from it.  Our
template spectra are listed in Table~1 and plotted with the corresponding
spectra of NGC~1052 in Figure~1.   Since there is very little stellar contribution to the
UV in elliptical galaxies and no evidence for hot stars in NGC~1052, no template was used for the G160L grating.

     We then measured the emission line fluxes using the subtracted spectrum
corrected for Galactic extinction.  Flux ratios relative to H$\beta$ are given
in Table~2.  To measure the line fluxes, uncontaminated continuum regions were
selected on either side of the features to be measured.  A linear fit was made
to the continuum underlying the emission features and the lines were extracted. 
In several cases, it was difficult to distinguish between real emission lines
and artifacts remaining from our attempt to remove the underlying spectum of the
host galaxy. This is most notable in the He~II~$\lambda$4686 emission. In
Table~2, we have indicated where these uncertainties are large. The emission line
widths are broader than typically observed in the narrow line regions of AGN,
corresponding to a range of velocities between
900~km~s$^{-1}$~$\leq$~FWHM~$\leq$~1100~km~s$^{-1}$ (by comparison, the
instrumental resolution for the high dispersion gratings is
FWHM$\sim$230~km~s$^{-1}$).

    In several cases, it was necessary to deblend line contributors before
accurate emission line fluxes could be derived.  In all cases, the spectra were
converted to velocity space to preserve the correct line widths.  For example,
H$\alpha$ is heavily blended with the [N~II] $\lambda\lambda$6548,6584 doublet. 
To extract the H$\alpha$ feature, the normalized line profile of H$\beta$ was
used to make a template for both of the [N~II] lines, since no broad wings were
detected on the H$\beta$ profile.  This technique proved more accurate than fitting
with Gaussian profiles because it approximates the actual structure of the
emission lines.  The [N~II] template was created by separating the two profiles
according to their intrinsic transition wavelengths and assuming the flux ratio
corresponds to that of their Einstein A-values
(A$_{\lambda6584}$/A$_{\lambda6548}$=2.96).  The wavelength offset and relative
flux of the template were then varied until the residual of the [N~II]
contribution subtracted from the H$\alpha$+[N~II] line blend gave the best fit
to H$\alpha$, namely, a single emission feature on a flat continuum (Figure 2a).
  Figure 2b shows the resulting H$\alpha$ fit compared to the H$\beta$ velocity
profile and demonstrates the reliability of this technique.

     Several other line blends were treated using the same technique. It is
important to deblend and provide accurate flux ratios for the
[S~II]~$\lambda\lambda$6716,6731 lines because they provide a low density
constraint for the emission line region.  To fit them, H$\beta$ profiles were
centered at the intrinsic wavelengths of the two [S~II] lines.  The flux ratio
was varied until residuals were minimized.  For line blends of other hydrogen
lines, namely H$\gamma$+[O~III]~$\lambda$4363 and
H$\epsilon$+[Ne~III]~$\lambda$3967, the H$\beta$ profile was used as a template
for the Balmer line since it will have an identical intrinsic shape.  All lines
which required deblending are noted in Table 2.

\subsubsection {Uncertainties}

     The primary sources of uncertainty are continuum placement in extracting
emission lines and the approximations associated with the deblending procedure. 
 For the first case, the uncertainty depends on the strength of the emission
line relative to the noise of the data and the existence of host galaxy
features.  We experimented with continuum levels for lines of different
strengths. For the optical
and near-UV data, where the host galaxy is dominant but spectral resolution is
high, this leads to an estimate of the uncertainty for $\lq\lq$strong'' lines
of  $\pm$5$\%$ (defined as having F~$\geq$~F$_{H\beta}$), for $\lq\lq$moderate
strength'' lines  $\pm$10$\%$ (0.2$\times$F$_{H\beta}$~$<$~F~$<$~F$_{H\beta}$), and for
$\lq\lq$weak'' lines $\pm$20$\%$ (F~$\leq$~0.2$\times$F$_{H\beta}$). For wavelengths below $\sim$2000~$\AA$, there is much less contribution
from the host galaxy, however, the signal-to-noise ratio is significantly
deteriorated.  Consequently, we find flux uncertainties of the moderate strength
lines to be $\pm$20$\%$ and the weak lines to be $\pm$30$\%$, for wavelengths
below 2000~$\AA$.   

      Additional uncertainties are introduced for lines that require deblending.
 To reliably estimate this uncertainty, the flux of the residual associated with
the [N~II]~$\lambda$6584 line (see Figure~2a) was measured and shown to be
$\pm$10$\%$ of the measured flux for the line.  This represents an upper limit
to the general deblending uncertainty because it was the largest relative
residual of all procedures.  These two uncertainties were then added in
quadrature to obtain the total error for the measured line ratios.  The total
uncertainty estimate for each emission line is listed in Table~2.

\subsubsection {Reddening}

    To estimate the internal reddening of NGC 1052, we compared our measured
H$\alpha$/H$\beta$ flux ratio (F$_{H\alpha}$/F$_{H\beta}$~=~4.53, corrected for
Galactic extinction) to an assumed intrinsic ratio (H$\alpha$/H$\beta$)$_I$. 
For N$_e$~=~10$^4$~cm$^{-3}$ and T$_e$~=~10,000~K and assuming case~B
recombination, (H$\alpha$/H$\beta$)$_I$~=~2.86, which is consistent with the
typical ratio determined for AGN (Osterbrock 1989).  Using the extinction curve
of Savage~\& Mathis (1979), this implies a reddening of E(B-V)~=~0.42~$\pm$.20, where
the uncertainty was derived from the errors associated with the H$\alpha$ and H$\beta$ flux
measurements. The
deduced intrinsic line ratios are listed in Table~2. Uncertainties were estimated from
propagation of errors associated with the reddening and observed flux ratios.  Since all previous studies
of NGC~1052 determined negligible internal reddening, it is important to examine
this issue in more detail (Koski~\& Osterbrock 1976; F78; Fosbury et~al.\ 1981; Ho et~al.\ 1993; Ho et~al.\ 1997).

    The intrinsic He~II~$\lambda$1640/He~II~$\lambda$4686 ratio is theoretically
a better measure of reddening than H$\alpha$/H$\beta$ because of the large span
in wavelength between the two lines (\cite{krae1994}).  However, the large
flux uncertainties associated with the He~II lines greatly diminish the
usefulness of this diagnostic for our data set.  Yet, we can still use it as a
check on our reddening deduced from the Balmer line ratio.  We assume an
intrinsic ratio of (He~II~$\lambda$1640/He~II~$\lambda$4686)$_I$~=~6.6,
corresponding to N$_e$~=~10$^4$~cm$^{-3}$ and T$_e$~=~10,000\arcdeg~K (\cite{seat1978}). 
This is consistent with the dereddened ratio of
He~II~$\lambda$1640/He~II~$\lambda$4686~$\geq$ 4.2 considering the large
uncertainties associated with these line fluxes (Table~2). 

    We can also use the relative flux ratios of the higher order hydrogen Balmer
lines beyond H$\beta$ as a consistency check.  A comparison of the dereddened
Balmer ratios with their intrinsic values is given in Table~3 (\cite{oste1989}).
They are seen to be consistent within the quoted uncertainties.  

     The colors of the host galaxy of NGC~1052, B-V~=~0.91, U-B~=~0.42,
are in accord with typical colors found in normal elliptical galaxies of similar
Hubble type, E4 (F78).  Thus, the large extinction found above is not
consistent if applied to the entire galaxy.  If the implied reddening is
correct, it suggests that dust is localized to the nuclear region of NGC~1052.  
There is evidence which supports this. The B-R and B-I images of Carter et~al.
(1983) and B-R image of Davies et~al.~(1986) clearly show a diffuse dust lane extending out 
from the central region to about
20$\arcsec$. Furthermore, the IR measurements of Becklin~et~al.~(1982) indicate
a large IR excess between 5-20~$\mu$m confined within 2$\arcsec$ of the nucleus.
 The strength of the IR flux at these wavelengths is not consistent with stellar
radiation emanating from cool stars and may be thermal radiation from heated
dust.

\subsection {H$\alpha$ Imagery}

     Images of the nuclear region of NGC~1052 were taken with WFPC2/{\it HST}
using the F658N filter on 1996~June~16.  These data were obtained from the STScI
data archives. This filter isolates any extended H$\alpha$ emission. In Figure~4, we have combined two exposures (with 800 and 900~s
exposure times) and removed cosmic ray events.  The size of the aperture used for 
the FOS observations is shown over the bright emission core for reference, 
although the precise position of the aperture is uncertain due to the guide star 
lock failure. Strong filamentary nebular H$\alpha$ emission is seen to extend about
1$\arcsec$ around the compact nucleus, with a more diffuse halo extending to
further distances.  At a position angle of $\sim$235$\arcdeg$ is a narrow filament of H$\alpha$
emission.  This is compared to the radio jet/lobe observed by Wrobel et~al.~(1984) at 1.5~GHz which is
centered at a position angle of $\sim$275$\arcdeg$.

\section{Photoionization Models}

    The purpose of the following analysis is to determine if the emission in the
 compact nuclear region of NGC~1052 is consistent with photoionization by a simple
 non-thermal continuum, characteristic of AGNs.  We seek the simplest model, making 
 the fewest number of assumptions, that gives a satisfactory fit to the dereddened
 intrinsic emission line spectrum. Discrepancies between our model results and the observed
spectrum then provide further insight into the actual physical conditions in NGC~1052.

     Photoionization models were computed using Version~90 of the computer code CLOUDY (\cite{ferl1996}).  
These models were parameterized by the gas density n$_H$, continuum shape, and ionization parameter

\begin{displaymath}
U = \frac{Q}{4\pi R^{2}n_{H}c}
\end{displaymath}

\noindent where Q is the number of hydrogen ionizing photons s$^{-1}$ and R the distance from the inner face of the cloud to the central continuum source.  When a good fit was achieved, observational constraints on the ionizing luminosity provided constraints on R.

     We adopted the following idealized properties of the emission line gas. In cases where more than one density component was required, we assumed that each component was photoionized by unattenuated radiation from the central source.  Thus we assume, {\it a~priori}, relatively small covering factors for any inner gas components.  Physically, this corresponds to a clumpy gas composed of radiation bounded clouds. We assume that the observed radiation is emitted from the $\lq\lq$front'' faces of the clouds, that is, the side of the clouds facing the central continuum source. Each component was assumed to have constant density. 
We began by omitting dust grains from the emission line gas, hence, in our initial 
model, we assume solar abundances and that all reddening occurs external to the line 
emitting region.  Thus, the abundances relative to hydrogen are:
He~=~1.00$\times$10$^{-1}$, C~=~3.55$\times$10$^{-4}$,  N~=~9.33$\times$10$^{-5}$, 
O~=~7.41$\times$10$^{-4}$, Ne~=~1.17$\times$10$^{-4}$, Mg~=~3.80$\times$10$^{-5}$, 
Si~=~3.55$\times$10$^{-5}$, S~=~1.62$\times$10$^{-5}$, and Fe~=~3.24$\times$10$^{-5}$ (\cite{grev1989}). 
The effects of grains mixed with the gas will be treated in
a later section.  

\subsection {Choice of Model Parameters:}

     We adopted a simple power law ionizing continuum parameterized 
by spectral index $\alpha$. Observations in the UV and X-ray were used to 
constrain its shape (Figure~3).  We assumed that the observed UV continuum flux was 
reddened by the same amount as the emission line fluxes, E(B-V)~=~0.42, as 
deduced from the H$\alpha$/H$\beta$ ratio.  This gives a projected intrinsic 
flux at the Lyman edge of 
F$_{1Ryd}$$\sim$1.9$\times$10$^{-27}$~ergs~s$^{-1}$~cm$^{-2}$~Hz$^{-1}$.  
Results from {\it ASCA} and {\it ROSAT} show NGC~1052 to have a rather flat spectrum 
between 2$-$6 keV, a steeper continuum between both 6$-$10 keV and $\sim$1$-$2~keV, 
and a flattening at energies below 1~keV (\cite{guai1999}).  This 
indicates that the 
soft X-ray emission is likely heavily absorbed, which is consistent with the
modeling by Weaver et~al.~(1998).  If this absorption is due to a dusty gas, then our previous assumption that the UV
continuum is attenuated by a column of dust is justified.  We chose the observed
{\it ASCA} flux 
at 6~keV, F$_{6 keV}$~=~1.4$\times$10$^{-30}$ ergs s$^{-1}$ cm$^{-2}$~Hz$^{-1}$, 
as the best estimate of the intrinsic nuclear X-ray flux, as shown in Figure~3.  Interpolating 
between the estimated intrinsic UV continuum flux at 1~Ryd and the X-ray flux at 6~keV gives $\alpha$~=~$-$1.2 and a total intrinsic luminosity for the ionizing continuum of 
L$_{> 1 Ryd}$$\sim$2.0$\times$10$^{42}$~ergs~s$^{-1}$. We note the large uncertainty in
this value due to the uncertainty in the amount of obscuration by dust.   Previous studies
have indicated that the continuum emission in AGNs is typically not reddened as much as the
emission line flux, thus, we may be overestimating the total luminosity (\cite{cost1977}).

     In Figure~3 we have plotted the spectral energy distribution of the
continuum emission from the radio through hard X-ray energies of the central point source.  
In each energy band, we have plotted only the emission that cannot be spatially resolved.  Data for both the observed and reddening corrected UV continuum are included,
illustrating the large effect that our
assumed extinction has on the energetics of the active nucleus.

     Simple model calculations using a gas with constant density do not match 
the intrinsic emission line spectrum of NGC~1052 for any choice of density.  One 
important problem with these models is that they cannot match both the auroral 
and nebular line fluxes simultaneously.  When a low density gas is used, 
n$_H$$\sim$10$^{3.5}$~cm$^{-3}$, the nebular lines are fit well (i.e., 
[O~III]~$\lambda$$\lambda$5007,4959, [N~II] $\lambda$$\lambda$6548,6584,  
[O~II]~$\lambda$3727, and [S~II] $\lambda$$\lambda$6716,6731), but their auroral line 
counterparts are under estimated, namely [O~III] $\lambda$4363, 
[O~II] $\lambda$2470, and [S~II] $\lambda$$\lambda$4069,4076.  For higher 
densities, the opposite is true.  As noted earlier, it was the high temperature 
implied by the [O~III]~$\lambda$4363/[O~III] $\lambda$$\lambda$4959,5007 ratio that led to 
the original interpretation of shock excitation for NGC~1052. Indeed, our 
[O~III] line flux ratio, [O~III]~$\lambda$4363/ [O~III] $\lambda\lambda$4959,5007$\sim$ 0.089, 
implies T$>$100,000~K in the low density limit (\cite{oste1989}), which is a factor $\sim$5 higher than we would expect in the O$^{++}$ zone of a photoionized nebulae.  However, choosing the high density that matches this ratio under-estimates the other nebular line fluxes because their critical 
densites are much lower than that of [O~III]~$\lambda$$\lambda$5007,4959, which have critical
density, n$_c$~=~6.5$\times$10$^5$~cm$^{-3}$.

    The detailed analysis of P84 showed this could be remedied by employing two 
separate gas components having different densities. In this scenario, the 
nebular lines overwhelm their auroral counterparts in a low density region 
where their ratios are determined solely by temperature.  Conversely, in a higher 
density gas the nebular line emissions are suppressed due to collisional 
de-excitation and, consequently, the auroral line fluxes become a more 
significant source of cooling.  Hence, a density stratification effect is 
mistaken for an indication of high temperature. Multi-component gases having a range
of densities are a common feature in models for LINERs and Seyfert galaxies (Halpern \& Steiner 1984; Filippenko 1985; Kraemer et~al. 1994, 1998).   Since we seek the simplest possible model, our method was to find the fewest number
of discrete cloud components that would reproduce the observed emission.  Although assuming a small number of distinct isolated gas components might be an oversimplification, it provides insight
into the physical conditions and structure of the emission line gas.  An alternate approach, the locally
optimally emitting cloud model, has also been shown to effectively model the narrow line region emission
in AGNs (\cite{ferg1997}).

     Thus, as a first modification to the simple model, we assumed a gas comprised of two density components. Our method was to fit the nebular lines that have very low critical 
densities ([S~II] $\lambda$$\lambda$6716,6731, 
[N~II] $\lambda$$\lambda$6548,6584, and [O~II] $\lambda$3727) with a low 
density gas, component~1, and the strongest auroral lines (ie, those of S$^+$, 
O$^+$, and O$^{+2}$) with a higher density gas, component~2.  Critical densities of various lines place constraints on the density for each component.  If the 
[O~II]~$\lambda$3727 line is a principal coolant for component~1, the gas density must not be significantly
greater than its critical density of 
n$_c$~=~4500~cm$^{-3}$ (\cite{dero1984}).  A lower limit on the 
density for component~1 is obtained from the [S~II]~$\lambda$$\lambda$6716,6731 doublet lines.  The flux ratio [S~II]~$\lambda$6716/ [S~II]~$\lambda$6731~=~1.1 indicates an electron density
n$_e$~$>$~300~cm$^{-3}$ (\cite{alle1984}).  For the high density component, the 
[O~III]~$\lambda$4363/ [O~III] $\lambda$5007 ratio implies 
6.5$\times$10$^5$~$\leq$~n$_e$~$\leq$ 3.3$\times$10$^7$ cm$^{-3}$ as determined by the 
critical densities of the two lines.  Additionally, the strong 
[O~I]~$\lambda$6300 emission indicates 
n$_e$~$\leq$~1.8$\times$10$^6$~cm$^{-3}$ (\cite{dero1984}).

      Since the spectrum of NGC~1052 is dominated by lines of low ionization species, 
lines emitted by S$^+$, N$^+$, O$^0$, O$^+$, and C$^+$, a low ionization parameter is required.  Figure~1 of Ferland~\& Netzer (1983), which
shows the relationship between various line strengths and the ionization parameter for a 
low density gas, indicates U~$<$~10$^{-3.5}$ for component~1.  We find that a slightly 
higher U is needed for component~2 to sufficiently populate the states responsible for the auroral line emission.  Model results using 
n$_H$~=~10$^{3.5}$~cm$^{-3}$, U$_1$~=~10$^{-3.8}$ and n$_H$~=~10$^{6.2}$~cm$^{-3}$, 
U$_2$~=~10$^{-3.1}$ for components one and two, respectively, are presented in 
columns~3 and~4 of Table~4.  The composite model in column~5 of Table~4 was created by assuming equal contribution from each component to the total observed 
H$\beta$ flux.  

\subsection {The Effects of Grains in the Emission Line Gas}

      Inspection of Table~4 shows that the overall intrinsic emission line
spectrum of NGC~1052 is fit very well by our photoionization model, which
has employed a minimum of assumptions.  Particulary, the strong auroral line 
fluxes of [O~III]~$\lambda$4363, [O~II]~$\lambda$2470, and [S~II] $\lambda\lambda$4069,4076
are matched well, as are the [O~III]~$\lambda$4363/[O~III]~$\lambda$5007 and
[O~II]~$\lambda$2470/ [O~II]~$\lambda$3727 ratios, which are often mistakenly
used as temperature indicators.  However, several of the prominent low ionization
features are considerably underpredicted.  For example, the observed
[N~II]~$\lambda\lambda$6548,6584, [S~II]~$\lambda\lambda$6716,6731, and 
[O~I]~$\lambda\lambda$6300,6363 line fluxes are matched within a factor of
two, but are all somewhat underestimated.  The 
C~II]~$\lambda$2326 fit is even worse.

     All of these poorly fit lines emit predominately in the warm
partially ionized zones, [S~II] $\lambda\lambda$6716, 6731 and [N~II]~$\lambda\lambda$6548,6584 most strongly in component~1 and [O~I] $\lambda\lambda$6300, 6363 and C~II]~$\lambda$2326 primarily in the high density component~2.  The characteristic emission
in these regions is likely greatly modified by the presence of dust due to the heavy depletion of important coolants (particularly magnesium and silicon) and an increase in gas temperature caused by the selective removal of photons near the Lyman
limit by dust grains.  To treat this, we have included grains typical of the Orion Nebula mixed with the emission line gas (\cite{bald1991}).  We selected this type of grain
structure since, of the available data for dust grains, physical conditions in 
an H~II region most closely approximate those of NGC~1052.  
Constraints on the dust-to-gas ratio in NGC~1052 come from the strengths of the Si~III]~$\lambda$$\lambda$1882,1892 and
Mg~II~$\lambda$2800 fluxes.  The calculated emission of these lines becomes too weak when the full
depletion deduced for Orion is used in our models (\cite{ferl1996}: based on results of \cite{bald1991},  \cite{rubi1991}, and \cite{oste1992}).  Therefore, we assumed half of the
depletion and grain abundances, giving the following elemental abundances relative to hydrogen:
He~=~9.77$\times$10$^{-2}$, C~=~3.31$\times$10$^{-4}$,  N~=~8.13$\times$10$^{-5}$, 
O~=~5.50$\times$10$^{-4}$, Ne~=~8.51$\times$10$^{-5}$, Mg~=~1.07$\times$10$^{-5}$, 
Si~=~1.20$\times$10$^{-5}$, S~=~1.29$\times$10$^{-5}$, and Fe~=~1.02$\times$10$^{-5}$.  

     Model results, using the same parameters as in our previous model are given in Table~5.   We have included model predictions of the [Ca~II] $\lambda\lambda$7291, 7324 lines for both the dusty and solar abundance models in Tables~5 and~4, respectively.  Ferland (1993) has shown that the weakness of these lines in the narrow line regions of some AGNs is a good indicator of the presence of grains - a sharp decrease in their strengths is expected with depletion.  Indeed, comparison of the results of our two models demonstrates this. Specifically, the dusty model predicts a decrease in the strength of the lines by a factor of ten.  Although the [Ca~II] 
lines lie outside of the spectral region covered by our data, the near infrared spectral data of Diaz et~al. (1985) clearly show that their strengths are overpredicted by our dust free model (see their Figure~1).  Thus, there is evidence that significant depletion exists in NGC~1052.  

     Our depletion percentage corresponds to a gas column-to-reddening scaling of roughly N$_H$/E(B-V)~=~1$\times$10$^{22}$~cm$^{-2}$~mag$^{-1}$ (\cite{knap1974}).  The ionized column densities of our gas components are comparatively small (N$_{HII}$~= 3.1$\times$10$^{19}$ cm$^{-2}$ and N$_{HII}$~=~3.6$\times$10$^{20}$~cm$^{-2}$, for components~1 and~2, respectively).  Thus, most of the reddening occurs in a neutral gas obscuring our line of sight to the 
emitting regions.  If this obscuration is due to gas component~1 of our model, n$_H$~= 10$^{3.5}$ cm$^{-3}$,  a cloud thickness of only $\sim\frac{1}{2}$~pc is required to produce the observed extinction. 

     The polarimetry study by Barth et~al.(1999) shows that [O~I]~$\lambda$6300 is more polarized than the other $\lq\lq$narrow'' emission lines between $\sim$4800$-$ 6800 $\AA$, with a polarization percentage more than twice as much as any other line.  This indicates that [O~I]~$\lambda$6300 may be transmitted through more dust than the other lines.  This effect is explained naturally, as in our model, if [O~I]~$\lambda$6300 is emitted predominately in an extended dusty partially ionized zone (PIZ).  If we are viewing the line emission primarily from the ionized faces of the gas clouds, then the lines emitted in the $\lq\lq$back side'' of the gas pass through a much larger dust column before escaping.  Our calculations show that the PIZ of component~2, where [O~I]~$\lambda$6300 is most strongly produced, has a much larger column density than the PIZ of component~1.  Since this is the region where the [N~II]~$\lambda$$\lambda$6548,6584 and [S~II]~$\lambda$$\lambda$6716,6731 lines are emitted, this explains why there is much less polarization of these lines than of the [O~I]~$\lambda$6300 line.

\subsection {Structure of the Emission Line Region}

     We now turn our attention toward the structure of the emission line region as deduced from our model results.  Since we have an estimate of the luminosity of ionizing radiation, we can calculate the radial distances of the different gas components from 
the putative central engine, R$_1$ and R$_2$, using U~$\propto$~L$_{(\geq1Ryd)}$/R$^2$.  
This gives R$_1$~$\sim$~100~pc and R$_2$~$\sim$~2~pc.  The deduced 
location of component~1 lies outside the projected aperture by about a 
factor of 2, however, the uncertainty in the intrinsic luminosity could account
for this discrepancy.  Furthermore, if the gas is spherically distributed, then much of it would be projected into the aperture. 

   The broad widths of the emission lines (900~$\leq$ FWHM $\leq$ 1100 km~s$^{-1}$) give clues to the gas dynamics in NGC~1052.  We find no correlation between line width and critical density.  Thus, since the high and low density components are at very different distances from the putative central massive object, the velocities of the two components cannot be explained as simple circular Keplerian orbits, for which v~$\propto$~r$^{-1/2}$. 
 
     As mentioned earlier, our assumption that the line emitting gas is comprised of distinct components having constant densities is a rough approximation, but qualitatively the 
higher density gas is seen to be closer to the center of the 
nucleus.  To match the calculated H$\beta$ luminosity with the observed luminosity, covering factors of $\epsilon$$_1$~=~0.28 and $\epsilon$$_2$~=~0.25 are required for components 
1 and 2, respectively, assuming that each component contributes equally
to the total H$\beta$ emission. Model predictions of L$_{H\beta}$ for each
component are given in Table~5.  Since the covering factors for each component were calculated independently of 
each other, the attenuation of the inner component was not treated and hence $\epsilon$$_1$ represents
a minimum covering factor for the outer component. Therefore, a minimum total covering factor for NGC~1052 is $\epsilon$$_T$~=~0.53, which represents the probability that the line-of-sight to the central continuum
is obscured by the emission line gas.  This is considerably higher than that determined for the narrow line regions of Seyfert~1 galaxies (\cite{netz1993}). Thus, our previous assumption that the central ionizing continuum is occulted by a dusty gas is vindicated.  This is consistent with the study by Barth et~al.~(1998) which concluded that many LINERs may have bright compact UV sources that are obscured by dust.

\subsection {The High Ionization Component}

     Although the spectrum of NGC~1052 is dominated by low ionization features,
there are several high ionization features present, namely, C~IV~$\lambda$1549, [Ne~IV]~$\lambda$2424, and [Ne~V]~$\lambda$3426.  As anticipated from our choice of low ionization parameters, U$_1$~\& U$_2$, these lines are greatly 
underestimated.  In particular, the predicted C~IV~$\lambda$1549 is $\sim$4 
times too weak and [Ne~IV]~$\lambda$2424 and 
[Ne~V]~$\lambda$3426 are underpredicted by a factor of $\sim$10.  Imposing the 
required U on either of the two gas components would severely degrade our fit 
to the prominent low ionization features, giving results characteristic of 
Seyfert galaxies.  Indeed, it is this lower ionization that distinguishes 
LINERs from other active galaxies. Thus, there is an indication that a highly 
ionized gas component is present in NGC~1052, although its contribution is small.

     The critical density of [Ne~IV]~$\lambda$2424 limits 
n$_e$$\leq$3$\times$10$^4$~cm$^{-3}$. However, for a gas with this low density and 
having a high ionization parameter, the [O~III]~$\lambda$$\lambda$5007,4959 
lines become the dominant coolants and their fluxes are far overpredicted.  
We find that an emitting gas that is matter bounded can account for this. For illustrative purposes, we have calculated the emission for a gas without dust grains having U$_3$~=~10$^{-1.5}$
and n$_H$~=~10$^{4.2}$~cm$^{-3}$ and a total column density of N$_H$~=~10$^{21}$~cm$^{-2}$.
This corresponds to a distance R$_3$$\sim$3~pc of the highly ionized gas from the
central continuum source. A very low covering factor of only $\sim$0.05 is required to match
the observed C~IV~$\lambda$1549 luminosity with this extra component, hence our assumption that
all components receive unattenuated continuum radiation remains valid.  This simple model for a highly ionized component matches the observed [Ne~V]~$\lambda$3426 emission line very well and the [Ne~IV]~$\lambda$2424 line within a factor of 2.  Furthermore, since the emissivity of C~IV~$\lambda$1549 is very high under these conditions, there would be only a small contribution from this component
and, therefore, no change in the low ionization line ratios predicted by a three-component model 
from those listed in Tables~4~\&~5.

\section{Discussion}

     We have shown that a simple photoionization model explains the full set of 
dereddened optical and UV emission lines observed in the FOS data of NGC~1052
quite well.   We note that the reddening correction we have applied to the UV continuum
is highly uncertain and that this enhanced luminosity is essential in reproducing the
line luminosities. Our best fit model employs a dusty emission line gas with a high
covering factor.  This is consistent with the apparent absorption of the soft
X-ray flux observed in the {\it ROSAT} data, indicating that the line-of-sight toward
the AGN is obscured. The flat intrinsic ionizing continuum
($\alpha$~=~$-$1.2) that we have deduced for NGC~1052 is in accord with the
recent study by Ho (1999) which demonstrated that a sample of LINERs all have a
higher X-ray/UV continuum ratio than found in typical high luminosity AGNs.

\subsection {Extended Emission Line Region}

     We examine if the deduced ionizing continuum flux is sufficient to
power the extended emission line region (EELR) emission described in
F78. They found a diffuse emission line gas extending out to about 20$\arcsec$ from the nucleus having a total Balmer line flux similar to that observed in the nuclear emission region.  Using their flux measurements and assuming that all of the emission line fluxes experience the extinction that we have deduced from our analysis gives L$_{H\beta}$$\sim$2.6$\times$10$^{40}$ ergs~s$^{-1}$ for the total H$\beta$ luminosity in the extended plus compact emission components.
A simple photon-counting argument (\cite{netz1993}) using our deduced continuum shape and Lyman edge flux demonstrates that a covering factor of $\sim$2 is required to produce the H$\beta$ luminosity.

     A covering factor greater than unity indicates that if all of the line emission in NGC~1052 (extended~+~compact) is powered by photoionization from the putative
non-thermal continuum, then our view of this continuum must be even more
obscured than we have previously deduced.
Furthermore, the large covering factor of the inner gas regions calculated in our model suggests
that the shape of the ionizing continuum incident upon the extended region would
be greatly modified.  However, the attenuation of UV continuum photons by
the emission line gas in the compact region would result in a much harder SED
incident upon the extended gas and a much higher actual covering factor for the EELR is
required.    An alternate explanation is that the central source emission is beamed
in NGC~1052 such that the continuum flux "seen" by the emitting gas in the extended regions
is enhanced.  Another possibility, is that some process
other than photoionization by the central source drives the emission in the EELR, either
stellar photoionization or 
shock excitation. A detailed analysis of the diffuse emission regions is needed to fully test this.  

\subsection {Effects of the Choice of Extinction Curve}

     As noted in section 3, one discrepancy in our model is the
underestimate of the C~II]~$\lambda$2326 flux by a factor of $\sim$2.  The
C~II]~$\lambda$2326 emission is strong in a gas with low ionization parameter
and high density. However, calculations show that the Mg~II $\lambda$2800 flux
is far overpredicted using these model conditions. Alternately, C~II]~$\lambda$2326 is produced predominantly in the PIZ of a photoionized gas, suggesting a stronger X-ray continuum
would strengthen the feature.  However, the [O~I]~$\lambda$$\lambda$6300,6363
line fluxes are more sensitive to the high energy flux and are greatly over-estimated if the X-ray continuum is enhanced. 
 
    A poor choice of the extinction curve may account for this discrepancy.  Several studies have shown that the UV dust extinction in extragalactic objects is not represented well by the Galactic Savage~\& Mathis curve (\cite{fitz1985,calz1994}).  Often a weaker 2200~$\AA$ absorption
feature and steeper rise in the far-UV are found. As an {\it ad~hoc} demonstration, we show how the UV line strengths are affected by removing the 2200~$\AA$ bump from the extinction curve in Table~6.  The [C~II]~$\lambda$2326 and [Ne~IV]~$\lambda$2424 line fluxes are both significantly reduced, illustrating that if an extinction law incorporating a strong 2200~$\AA$ feature were erroneously applied, it could lead to a substantial overestimate of the intrinsic strengths of these lines.

\section{Conclusions}

      We have analyzed archival {\it HST}/FOS data of the prototypical LINER galaxy NGC~1052.  The aperture sampled the inner $\sim$~82~pc$\times$82~pc of the nuclear region, thereby excluding the extended emission region.  One important result is that the deduced H$\alpha$/H$\beta$ ratio provides evidence for considerable extinction (E(B-V) =0.42) of the emission line fluxes in NGC~1052.  Dust appears to be concentrated in or near the nucleus and may also be attenuating the nonthermal continuum of a compact source, which supports the Barth et~al.~(1998) conclusion that the nuclei of many LINERs are obscured by dust. 
 
     From our modeling results, we find that the nuclear emission observed in the inner region of NGC~1052 is explained well by photoionization from a simple power law continuum having a flat spectral index, $\alpha$=$-$1.2.  Moreover, the ionizing continuum is consistent with our deduced reddening and observed flux distribution. This is also consistent with the study by Ho (1999) which showed that a sample of LINERs all have a flat spectral energy distribution in the ionizing continuum.

     Most of the dereddened line fluxes are matched by a model having two dusty gas components of different densities, photoionized by a single power law continuum.  The appearance of the C~IV~$\lambda$1549, [Ne~IV]~$\lambda$2424, and [Ne~V]~$\lambda$3426 lines in the spectrum indicates a third gas component may be present, characterized by a high ionization parameter.  The key exception to our good model fit is the prominent C~II]~$\lambda$2326 line, which is underpredicted by a factor of about two. This flux is heavily affected by the strength of the 2200~$\AA$ absorption feature. Hence, the C~II]~$\lambda$2326 line would be overestimated if the 2200~$\AA$ bump of the Savage \& Mathis extinction curve is not applicable to NGC~1052. This would also help explain the underestimate of the [Ne~IV]~$\lambda$2424 flux.

      The large widths of the emission lines and lack of correlation between width and critical density suggests non-gravitational motion of the emission line gas. 
 
     We conclude that a pure central source photoionization model with the simplest non-thermal continuum (a simple power law) reproduce the emission line fluxes in the inner region of NGC~1052 quite well.  Other processes
such as shocks or photoionization by stars are not required to produce the observed emission. However, we
cannot rule out the possibility that these mechanisms contribute to the extended nebular emission.

\acknowledgments

     This research has made use of the SIMBAD database operated at CDS, Strasbourg, France and of the NASA/IPAC Extragalactic Database (NED) which is operated by the Jet Propulsion Laboratory, California Institute of Technology, under contract with the National Aeronautics and Space Administration.  We also would like to acknowledge support for the research from NASA grants
NAG5-3378 to the IACS at The Catholic University of America.  Also, facilities of the Laboratory
for Astronomy and Solar Physics at NASA/GSFC were used in performing part of the research.  Finally, we are grateful to the referee, Gary Ferland, for his instructive comments and suggestions.

\clearpage

\figcaption[Fig1.eps]{FOS spectra of the nucleus of NGC~1052 from 1250 to 6800~$\AA$, corrected 
for Galactic reddening.  The plotted template spectra are scaled to match the absorption features 
of NGC~1052 and offset in flux for clarity.  }

\figcaption[Fig2.eps]{ (2a) Deblending procedure for H$\alpha$+[N~II].  The solid line is the H$\alpha$+[N~II] line blend with the host galaxy template removed.  The [N~II]~$\lambda\lambda$6548,6584 emission line template (dot-dashed line) was created from the H$\beta$ profile and subtracted to give the H$\alpha$ (dashed line) emission line.
(2B) Comparison of the shape of the deblended H$\alpha$ profile to the H$\beta$ profile scaled to match in flux. }

\figcaption[Fig3.eps]{  The multiwavelength spectrum of the nucleus of NGC~1052. The radio data (+) are the measurements of the compact emission ($\leq$~1$\arcsec$) by Wrobel et~al. 1984. The IR data points ($\ast$) are within $<$~2$\arcsec$ (Becklin et~al. 1982). The observed and dereddened UV continuum data at 1300~$\AA$ are from this study, $\sim$0$\arcsec$.86 ($\diamond$).  {\it ASCA} and {\it ROSAT} data ($\triangle$) are from Guianazzi \& Antonelli (1999) with a resolution of ~5$\arcmin$. It is unclear how much of this X-ray emission is compact.  The ionizing continuum used in
our photoionization models is shown as the solid line.  The dashed line is the attenuated continuum, assuming the central engine is viewed through the high density gas, component~2.  }

\figcaption[Fig4.ps]{  WFPC2 F658N image of the inner region of NGC~1052. The size of the FOS 0.$\arcsec$86 square aperture is plotted over the compact emission core for reference. North is up in this image.}

\clearpage

\begin{deluxetable}{lrrrr}
\tablewidth{0pt}
\tablecaption{FOS Observations of NGC~1052}
\tablehead{
\colhead{Data Set} & \colhead{Grating} &
\colhead{Resolution } & \colhead{Exposure } &
\colhead{Template Galaxy} \\
\colhead{} & \colhead{} &
\colhead{(\AA)} & \colhead{(s)} &
\colhead{} }
\startdata
Y3K80105T  &  G160L\tablenotemark{a} & $\sim$6.4\tablenotemark{b} & 1390 &
\nodata \nl
Y3K80106T  &  G160L &  & 2430 &   \nl
Y3K80107T  &  G160L &  & 2430 &   \nl
Y3K80108T & G270H     & $\sim$2.1\tablenotemark{c} & 2320 &  NGC~3608 \nl
Y3K80109T  & G400H     & $\sim$3.1\tablenotemark{c} & 1200 & NGC~3608 \nl
Y3K8010AT & G570H     & $\sim$4.4\tablenotemark{c} & 890 & NGC~5845 \nl
\enddata

\tablenotetext{a}{G160L spectra coadded, weighted by exposure time}
\tablenotetext{b}{$\lambda$/$\Delta$$\lambda$$\sim$250}
\tablenotetext{c}{$\lambda$/$\Delta$$\lambda$$\sim$1300}
 
\end{deluxetable}

\clearpage
\begin{deluxetable}{lrr}
\tablewidth{0pt}
\tablecaption{Measured and Dereddened Emission Line Ratios\tablenotemark{a}}
\tablehead{
\colhead{} & \colhead{E(B-V)=0.02} &
\colhead{E(B-V)=0.44}} 
\startdata
C IV $\lambda$1549 & 	0.14 ($\pm$ 0.04) & 0.75 ($\pm$ 0.64) \nl 
He II $\lambda$1640 & 	0.16 ($\pm$ 0.05) & 0.80 ($\pm$ 0.66) \nl
[O III] $\lambda$1663 & 0.07 ($\pm$ 0.02) & 0.35 ($\pm$ 0.28) \nl
N III] $\lambda$1750 & 	0.04 ($\pm$ 0.01) & 0.20 ($\pm$ 0.16) \nl
Si II + [Ne III] $\lambda$1808-1820  & 	0.10 ($\pm$ 0.03)	 &  \nl
Si III] $\lambda$$\lambda$1882,1892\tablenotemark{b} & 0.14 ($\pm$ 0.05) &
 0.80 ($\pm$ 0.71) \nl
C III] $\lambda$1909\tablenotemark{b} & 0.35 ($\pm$ 0.08) & 2.00 ($\pm$ 1.73) \nl 
N II] $\lambda$2140  & 	0.06 ($\pm$ 0.01)	 &  \nl
C II] $\lambda$2326\tablenotemark{c} & 0.58 ($\pm$ 0.08) & 3.80 ($\pm$ 3.51) \nl
[Ne IV] $\lambda$2424 & 0.02 ($\pm$ 0.01) & 0.10 ($\pm$ 0.09) \nl
[O II] $\lambda$2470 & 	0.14 ($\pm$ 0.03) & 0.60 ($\pm$ 0.43) \nl
Fe II $\lambda$2558-2583 & 	0.03 ($\pm$ 0.01)	 &   \nl
Fe II $\lambda$2601-2640 & 	0.14 ($\pm$ 0.03)	 &   \nl
Fe II $\lambda$2660-2684 & 	0.04 ($\pm$ 0.01)	 &   \nl
Fe II $\lambda$2711-2765 & 	0.15 ($\pm$ 0.03)	 &   \nl
Fe II $\lambda$2823-2844 & 	0.04 ($\pm$ 0.01)	 &  \nl
Mg II $\lambda$2800 & 	::0.46 \tablenotemark{d}  & 1.2 \nl
He II $\lambda$3204 & 	::0.05  & ::0.08  \nl
[Ne V] $\lambda$3346 & 	::0.015 & ::0.03 	  \nl
[Ne V] $\lambda$3426 & 	0.07 ($\pm$ 0.02) & 0.11 ($\pm$ 0.04) \nl
[O II] $\lambda$3727 & 	2.77 ($\pm$ 0.31) & 3.98 ($\pm$ 0.81) \nl
[Ne III] $\lambda$3869 & 0.60 ($\pm$ 0.08) & 0.82 ($\pm$ 0.16) \nl
[Ne III] $\lambda$3967\tablenotemark{b} & 0.20 ($\pm$ 0.03) & 0.27 ($\pm$ 0.05) \nl
H$\epsilon$ $\lambda$3970\tablenotemark{b} & 0.14 ($\pm$ 0.03) & 0.19 
($\pm$ 0.05) \nl
[S II] $\lambda$$\lambda$4069,4076 & 	0.68 ($\pm$ 0.10) & 0.88 ($\pm$ 0.17) \nl
H$\delta$ $\lambda$4102 & 0.24 ($\pm$ 0.03) & 0.31 ($\pm$ 0.05) \nl
H$\gamma$ $\lambda$4340\tablenotemark{b} & 0.44 ($\pm$ 0.08) & 0.53 ($\pm$ 0.11) \nl
[O III] $\lambda$4363\tablenotemark{b} & 0.27 ($\pm$ 0.05) & 0.32 ($\pm$ 0.06) \nl
He II $\lambda$4686 & ::0.18 & ::0.19 \nl
H$\beta$ $\lambda$4861 & 1.00\tablenotemark{e} & 1.00 \tablenotemark{f} \nl
[O III] $\lambda$4959 & 0.91 ($\pm$ 0.13) & 	0.88 ($\pm$ 0.13)\nl
[O III] $\lambda$5007 & 2.85 ($\pm$ 0.32) & 	2.72 ($\pm$ 0.31) \nl
[N I]  $\lambda$5200 & 0.23 ($\pm$ 0.03)  &  0.21 ($\pm$ 0.03) \nl
[N II] $\lambda$5755 & 	::0.09  & ::0.07 \nl
He I $\lambda$5876 & ::0.14  & ::0.10  \nl
[O I] $\lambda$6300 & 3.32 ($\pm$ 0.37)	 & 2.22 ($\pm$ 0.49) \nl
[O I] $\lambda$6364 & 1.13 ($\pm$ 0.13)	 & 0.74 ($\pm$ 0.17) \nl
[N II] $\lambda$6548\tablenotemark{b} & 1.73 ($\pm$ 0.26) & 1.10 ($\pm$ 0.29) \nl
H$\alpha$ $\lambda$6563\tablenotemark{b} & 4.53 ($\pm$ 0.68) & 2.86 ($\pm$ 0.77) \nl
[N II] $\lambda$6583\tablenotemark{b} & 5.14 ($\pm$ 0.77) & 3.24 ($\pm$ 0.87) \nl
He I $\lambda$6678 & ::0.07  & 	::0.04\nl
[S II] $\lambda$6716\tablenotemark{b} & 2.56 ($\pm$ 0.38) & 1.57 ($\pm$ 0.43)\nl
[S II] $\lambda$6731\tablenotemark{b} & 2.35 ($\pm$ 0.35) & 1.45 ($\pm$ 0.40) \nl
\enddata

\tablenotetext{a}{Relative to H$\beta$}
\tablenotetext{b}{Line deblending procedure required for flux measurement}
\tablenotetext{c}{Contains [O III] $\lambda$2321 and Si II] $\lambda$2335
emission}
\tablenotetext{d}{Mg II $\lambda$2800 is heavily absorbed; a fit to the doublet with
Gaussian features gives Mg~II~$\lambda$2800/H$\beta$~=~0.80}
\tablenotetext{e}{F$_{H\beta}$ =
2.65($\pm$0.13)$\times$10$^{-14}$~ergs~s$^{-1}$~cm$^{-2}$}
\tablenotetext{f}{F$_{H\beta}$~=~1.1($\pm$0.7)$\times$10$^{-13}$~ergs~s$^{-1}$~cm$^{-2}$}

\end{deluxetable}

\clearpage
\begin{deluxetable}{lrr}
\tablewidth{0pt}
\tablecaption{Intrinsic and Dereddened Balmer Line Ratios\tablenotemark{a}}
\tablehead{
\colhead{} & \colhead{Intrinsic\tablenotemark{b}} &
\colhead{Dereddened\tablenotemark{c}} }
\startdata 	
H$\alpha$ $\lambda$6563	& 2.86  &  2.86 ($\pm$ 0.77) \nl
H$\gamma$ $\lambda$4340 & 0.47  &  0.53 ($\pm$ 0.11) \nl
H$\delta$ $\lambda$4102 & 0.26  &  0.31 ($\pm$ 0.05) \nl
H$\epsilon$ $\lambda$3970 & 0.16  &  0.19 ($\pm$ 0.05) \nl
\enddata

\tablenotetext{a}{Relative to H$\beta$}
\tablenotetext{b}{From Osterbrock 1989}
\tablenotetext{c}{E(B-V)$_T$=E(B-V)$_G$+E(B-V)$_I$=0.44}

\end{deluxetable}

\clearpage

\begin{deluxetable}{lrrrr}
\tablewidth{0pt}
\tablecaption{Line Ratios From Model without Dust\tablenotemark{a}}
\tablehead{
\colhead{} & \colhead{Observed\tablenotemark{b}} &
\colhead{Comp 1\tablenotemark{c}} & \colhead{Comp 2\tablenotemark{d}} &
\colhead{Composite\tablenotemark{e}} } 
\startdata
C IV $\lambda$1549 & 0.75 ($\pm$ 0.64) & - & 0.28 & 0.14 \nl 
He II $\lambda$1640 & 0.80 ($\pm$ 0.66) & 1.09 & 1.28 & 1.19 \nl
[O III] $\lambda$1663 & 0.35 ($\pm$ 0.28)  & - & 0.49 & 0.25  \nl
N III] $\lambda$1750 & 0.20 ($\pm$ 0.16)  & - & 0.20 & 0.10 \nl
Si III] $\lambda$1882,1892 & 0.80 ($\pm$ 0.71) & 0.01 & 1.12 & 0.57  \nl
C III] $\lambda$1909 & 2.00 ($\pm$ 1.73)  &  0.21 & 3.50 & 1.86  \nl
C II] $\lambda$2326\tablenotemark{g}  & 3.80 ($\pm$ 3.51)  & 1.00 & 2.56 & 1.78 \nl
[Ne IV] $\lambda$2424  & 0.10 ($\pm$ 0.09) & - & 0.01 & 0.01  \nl
[O II] $\lambda$2470 & 0.60 ($\pm$ 0.43)  & 0.30 & 1.11 & 0.71  \nl
Mg II $\lambda$2800  & ::1.2 & 1.77 & 4.50 & 3.14  \nl
He II $\lambda$3204 & ::0.08   & 0.07 & 0.08 & 0.08 \nl
[Ne V] $\lambda$3346  & ::0.03  & - & - & $<$0.01  \nl
[Ne V] $\lambda$3426 & 0.11 ($\pm$ 0.04)  &  -	& - &  $<$0.01 \nl
[O II] $\lambda$3727 & 3.98 ($\pm$ 0.81)  & 7.07 & 0.09 & 3.58 \nl
[Ne III] $\lambda$3869  & 0.82 ($\pm$ 0.16)  & 0.82 & 1.92 & 1.37  \nl
[Ne III] $\lambda$3967 & 0.27 ($\pm$ 0.05) & 0.25 & 0.60 & 0.43 \nl
H$\epsilon$ $\lambda$3970 & 0.19 ($\pm$ 0.05)  & 0.17 & 0.16 & 0.17  \nl
[S II] $\lambda$$\lambda$4069,4076 & 0.88 ($\pm$ 0.17) & 0.49 & 0.92 & 0.71 \nl
H$\delta$ $\lambda$4102 & 0.31 ($\pm$ 0.05)  & 0.27 & 0.27 & 0.27  \nl
H$\gamma$ $\lambda$4340 & 0.53 ($\pm$ 0.11)  & 0.47 & 0.47 & 0.47  \nl
[O III] $\lambda$4363  & 0.32 ($\pm$ 0.06) & - & 0.59 & 0.30  \nl
He II $\lambda$4686 & ::0.19  & 0.12 & 0.14 & 0.13  \nl
H$\beta$  & 1.00 & 1.00 & 1.00	& 1.00 \nl
[O III] $\lambda$4959 & 0.88 ($\pm$ 0.13) & 0.34 & 1.72 & 1.03   \nl
[O III] $\lambda$5007 & 2.72 ($\pm$ 0.32)& 0.97 & 4.98 & 2.98  \nl
[N I]  $\lambda$5200 & 0.21 ($\pm$ 0.03) & 0.25 & 0.01 & 0.13  \nl
[N II] $\lambda$5755 &  ::0.07 & 0.05 & 0.17 & 0.11  \nl
He I $\lambda$5876 & ::0.10  & 0.17 & 0.15 & 0.16  \nl
[O I] $\lambda$6300 & 2.22 ($\pm$ 0.49)  & 1.23 & 2.19 & 1.71   \nl
[O I] $\lambda$6364 & 0.74 ($\pm$ 0.17)  & 0.39 & 0.70 & 0.55   \nl
[N II] $\lambda$6548  & 1.10 ($\pm$ 0.29) & 1.36 & 0.29 & 0.83   \nl
H$\alpha$ $\lambda$6563  & 2.86 ($\pm$ 0.77) & 2.92 & 2.96 & 2.94  \nl
[N II] $\lambda$6583  & 3.24 ($\pm$ 0.87) & 4.03 & 0.85 & 2.44  \nl
He I $\lambda$6678 &    ::0.04 & 0.05 & 0.03 & 0.04   \nl
[S II] $\lambda$6716 & 1.57 ($\pm$ 0.43) & 1.47 & 0.06 & 0.77   \nl
[S II]$\lambda$6731  & 1.45 ($\pm$ 0.40) & 2.03 & 0.14 & 1.09  \nl
[Ca II]$\lambda$7291  & -  & 0.61 & 0.06 & 0.34  \nl
[Ca II]$\lambda$7324  & -  & 0.40 & 0.04 & 0.22  \nl
\enddata

\tablenotetext{a}{Relative to H$\beta$}
\tablenotetext{b}{Corrected for Galactic and internal reddening, E(B-V)=0.44}
\tablenotetext{c}{U=10$^{-3.8}$, N$_H$=10$^{3.5}$}
\tablenotetext{d}{U=10$^{-3.1}$, N$_H$=10$^{6.2}$}
\tablenotetext{e}{Composite of components 1 and 2, assuming equal contribution to H$\beta$ flux}
\tablenotetext{g}{Contains [O III] $\lambda$2321 and Si II] $\lambda$2335 emission}

\end{deluxetable}

\clearpage
\begin{deluxetable}{lrrrr}
\tablewidth{0pt}
\tablecaption{Best-Fit Model:Includes Dust Grains \tablenotemark{a}}
\tablehead{
\colhead{} & \colhead{Observed\tablenotemark{b}} &
\colhead{Comp 1\tablenotemark{c}} & \colhead{Comp 2\tablenotemark{d}} &
\colhead{Composite\tablenotemark{e}} } 
\startdata
C IV $\lambda$1549 & 0.75 ($\pm$ 0.64) & - & 0.38 & 0.19 \nl 
He II $\lambda$1640 & 0.80 ($\pm$ 0.66) & 0.84 & 1.28 & 1.06 \nl
[O III] $\lambda$1663 & 0.35 ($\pm$ 0.28)  & 0.01 & 0.53 & 0.27  \nl
N III] $\lambda$1750 & 0.20 ($\pm$ 0.16)  & - & 0.25 & 0.13 \nl
Si III] $\lambda$$\lambda$1882,1892 & 0.80 ($\pm$ 0.71) & 0.02 & 0.54 & 0.28  \nl
C III] $\lambda$1909 & 2.00 ($\pm$ 1.73)  &  0.35 & 4.53 & 2.44  \nl
C II] $\lambda$2326\tablenotemark{g}  & 3.80 ($\pm$ 3.51)  & 1.45 & 2.86 & 2.16 \nl
[Ne IV] $\lambda$2424  & 0.10 ($\pm$ 0.09) & - & 0.01 & 0.01  \nl
[O II] $\lambda$2470 & 0.60 ($\pm$ 0.43)  & 0.36 & 1.02 & 0.69  \nl
Mg II $\lambda$2800  & ::1.2 & 0.89 & 1.16 & 1.03  \nl
He II $\lambda$3204 & ::0.08   & 0.05 & 0.08 & 0.07 \nl
[Ne V] $\lambda$3346  & ::0.03  & - & - & $<$0.01  \nl
[Ne V] $\lambda$3426 & 0.11 ($\pm$ 0.04)  &  -	& - &  $<$0.01 \nl
[O II] $\lambda$3727 & 3.98 ($\pm$ 0.81)  & 7.35 & 0.09 & 3.72 \nl
[Ne III] $\lambda$3869  & 0.82 ($\pm$ 0.16)  & 0.78 & 1.68 & 1.23  \nl
[Ne III] $\lambda$3967 & 0.27 ($\pm$ 0.05) & 0.23 & 0.51 & 0.37 \nl
H$\epsilon$ $\lambda$3970 & 0.19 ($\pm$ 0.05)  & 0.17 & 0.16 & 0.16  \nl
[S II] $\lambda$$\lambda$4069,4076 & 0.88 ($\pm$ 0.17) & 0.55 & 1.06 & 0.81 \nl
H$\delta$ $\lambda$4102 & 0.31 ($\pm$ 0.05)  & 0.26 & 0.27 & 0.27  \nl
H$\gamma$ $\lambda$4340 & 0.53 ($\pm$ 0.11)  & 0.47 & 0.47 & 0.47  \nl
[O III] $\lambda$4363  & 0.32 ($\pm$ 0.06) & 0.01 & 0.58 & 0.30  \nl
He II $\lambda$4686 & ::0.19  & 0.09 & 0.14 & 0.12  \nl
H$\beta$\tablenotemark{h} & 1.00 & 1.00 & 1.00	& 1.00 \nl
[O III] $\lambda$4959 & 0.88 ($\pm$ 0.13) & 0.33 & 1.50 & 0.92   \nl
[O III] $\lambda$5007 & 2.72 ($\pm$ 0.31)& 0.96 & 4.34 & 2.45   \nl
[N I]  $\lambda$5200 & 0.21 ($\pm$ 0.03) & 0.40 & 0.02 & 0.21  \nl
[N II] $\lambda$5755 &  ::0.07 & 0.06 & 0.19 & 0.13    \nl
He I $\lambda$5876 & ::0.10  & 0.16 & 0.14 & 0.15  \nl
[O I] $\lambda$6300 & 2.22 ($\pm$ 0.49)  & 1.40 & 2.49 & 1.95   \nl
[O I] $\lambda$6364 & 0.74 ($\pm$ 0.17)  & 0.45 & 0.79 & 0.62   \nl
[N II] $\lambda$6548  & 1.10 ($\pm$ 0.29) & 1.44 & 0.36 & 0.90   \nl
H$\alpha$ $\lambda$6563  & 2.86 ($\pm$ 0.77) & 2.93 & 2.92 & 2.93  \nl
[N II] $\lambda$6583  & 3.24 ($\pm$ 0.87) & 4.26 & 1.05 & 2.66  \nl
He I $\lambda$6678 &    ::0.04 & 0.04 & 0.03 & 0.04   \nl
[S II] $\lambda$6716 & 1.57 ($\pm$ 0.43) & 1.72 & 0.10 & 0.91   \nl
[S II]$\lambda$6731  & 1.45 ($\pm$ 0.40) & 2.20 & 0.23 & 1.22  \nl
[Ca II]$\lambda$7291  & -  & 0.07 & 0.01 & 0.04  \nl
[Ca II]$\lambda$7324  & -  & 0.04 & 0.01 & 0.03  \nl
\enddata

\tablenotetext{a}{Relative to H$\beta$}
\tablenotetext{b}{Corrected for Galactic and internal reddening, E(B-V)=0.44}
\tablenotetext{c}{U=10$^{-3.8}$, N$_H$=10$^{3.5}$}
\tablenotetext{d}{U=10$^{-3.1}$, N$_H$=10$^{6.2}$}
\tablenotetext{e}{Composite of components 1 and 2, assuming equal contribution to H$\beta$ flux}
\tablenotetext{g}{Contains [O~III]~$\lambda$2321 and Si~II]~$\lambda$2335 emission}
\tablenotetext{h}{L$_{H\beta}^{Obs}$=5.0$\times$10$^{39}$ ergs s$^{-1}$;
L$_{H\beta}^{Comp1}$=9.1$\times$10$^{39}$ ergs s$^{-1}$; 
L$_{H\beta}^{Comp2}$=1.0$\times$10$^{40}$ ergs s$^{-1}$ }
\end{deluxetable}

\clearpage
\begin{deluxetable}{lrrr}
\tablewidth{0pt}
\tablecaption{The Effect of the 2200~$\AA$ Feature in the Extinction Curve}
\tablehead{
\colhead{} & \colhead{S \& M Curve\tablenotemark{a}} &
\colhead{Modified S \& M Curve\tablenotemark{a},\tablenotemark{b}} & 
\colhead{Best-Fit Model}}
\startdata
C IV $\lambda$1549 &  0.75 ($\pm$ 0.64) & 0.75  & 0.19 \nl
He II $\lambda$1640 & 	0.80 ($\pm$ 0.66)  & 0.80  & 1.06 \nl
[O III] $\lambda$1663 &  0.35 ($\pm$ 0.28)  & 0.35  & 0.27 \nl
N III] $\lambda$1750 & 	0.20 ($\pm$ 0.16)  & 0.20  & 0.13 \nl
Si III] $\lambda$$\lambda$1882,1892 & 0.80 ($\pm$ 0.71) & 0.70  & 0.28 \nl
C III] $\lambda$1909 & 	 2.00 ($\pm$ 1.73)  &  1.69  & 2.44 \nl
C II] $\lambda$2326\tablenotemark{c} & 3.80 ($\pm$ 3.51) & 2.01 & 2.16 \nl
[Ne IV] $\lambda$2424 &  0.10 ($\pm$ 0.09) &  0.07 & 0.01  \nl
[O II] $\lambda$2470 & 	0.60 ($\pm$ 0.43)  & 0.39  &  0.69  \nl
\enddata

\tablenotetext{a}{E(B-V)$_T$~=~0.44}
\tablenotetext{b}{Savage \& Mathis extinction curve with 2200~$\AA$ feature
removed, E(B-V)$_T$~=~0.44}
\tablenotetext{c}{Contains [O III] $\lambda$2321 and Si~II]~$\lambda$2335
emission}

\end{deluxetable}

\end{document}